\documentclass{llncs}
\usepackage[utf8]{inputenc}

\usepackage{graphicx}
\usepackage{phdalgo}
\usepackage{color}
\usepackage{array}
\usepackage{xspace} 
\usepackage{caption}
\usepackage{subcaption}
\usepackage{hyperref} 
\usepackage{verbatim}
\usepackage{cite}
\usepackage{paralist}
\usepackage{amsmath, mathtools}
\usepackage{tikz}
\usetikzlibrary{shapes,arrows,automata,decorations,trees, backgrounds, patterns, calc}

\usepackage{xcolor}
\definecolor{darkgreen}{rgb}{0.0 0.5 0.0}
\definecolor{whitegreen}{rgb}{0.0 0.75 0.0}
\definecolor{whiteblue}{rgb}{0.0 0.0 1.2}
\definecolor{darkred}{rgb}{0.8 0.0 0.0}

\newcommand{\N}{\mathbb{N}}					% L'ensemble des entiers naturels
\newcommand{\R}{\mathbb{R}}					% Le corps des nombres rels
\newcommand{\cB}{\mathcal{B}}
\newcommand{\cF}{\mathcal{F}}
\newcommand{\etc}{\textit{etc}}
\newcommand{\ie}{\textit{i.e.}}

\newcommand{\fsa}{f_{\text{sa}}}%stephane f_{sa} reads f_{s \times a}, spacing bewteen s and a is not nice
\newcommand{\fpop}{f_{\text{pop}}}
\newcommand{\Kpop}{K_{\text{pop}}}
\newcommand{\Xopt}{\mathbf{x}_\mathit{opt}}
\newcommand{\Xsdp}{\mathbf{x}_\mathit{sdp}}
\newcommand{\bop}{\mathtt{bop}}
\newcommand{\nbb}{\#{\text{boxes}}}

\newcommand{\Ksa}{K_{\mbox{\scriptsize sa}}}
\renewcommand{\geq}{\geqslant}
\renewcommand{\leq}{\leqslant}

\usepackage{amsmath}
\usepackage{amsfonts}
\usepackage{amssymb} 
\usepackage{latexsym}
\usepackage{multirow}
\numberwithin{equation}{section}

\DeclareMathOperator{\parab}{par}

\newcommand{\gradf}[2]{\mathcal{D}(#1) (#2)}
\newcommand{\hessf}[2]{\mathcal{D}^2 (#1) (#2)}

\newcommand{\sampapprox}[4]{\mathtt{template\_optim} (#1, #2, #3, #4)}
\newcommand{\buildtemplate}[6]{\mathtt{build\_template} (#1, #2, #3, #4, #5, #6)}
\newcommand{\tmalgo}[2]{\mathtt{build\_quadratic\_form} (#1, #2)}
\newcommand{\lift}{n_{\text{lifting}}}
\newcommand{\nt}{n_{\mathcal{T}}}
\newcommand{\templateoptim}{\mathtt{template\_optim}}
\newcommand{\iasos}{\mathtt{ia\_sos}}
\newcommand{\minsa}{\mathtt{min\_sa}}
\newcommand{\maxsa}{\mathtt{max\_sa}}
\newcommand{\xb}{\mathbf{x}} 
\newcommand{\yb}{\mathbf{y}} 
\newcommand{\zb}{\mathbf{z}}
\begin{document}

\mainmatter
\title{Certification of Bounds of Non-linear Functions: the Templates Method}
%Efficient
%The Templates method in formal global optimization: efficient and safe lower bounds computation for non-linear programming}
%\title{The Templates method: efficient formal verification of non-linear programs}
% programs or problems?
%\title{Efficient Certification of Bounds of Non-linear Problems with Max-plus Templates and SDP}% ce titre donne d'abord le problème puis la solution

\author{Xavier Allamigeon\inst{1} \and St\'{e}phane Gaubert\inst{2} \and Victor Magron\inst{3}\and Benjamin Werner\inst{4}}

\institute{INRIA and CMAP, \'Ecole Polytechnique, Palaiseau, France,
\\
\email{Xavier.Allamigeon@inria.fr}
\and
INRIA and CMAP, \'Ecole Polytechnique, Palaiseau, France,
\\
\email{Stephane.Gaubert@inria.fr}
\and
INRIA  and LIX, \'Ecole Polytechnique, Palaiseau, France,
\\
\email{magron@lix.polytechnique.fr}
\and
INRIA  and LIX, \'Ecole Polytechnique, Palaiseau, France,
\\
\email{benjamin.werner@polytechnique.edu}
}

\maketitle

\begin{abstract}
  The aim of this work is to certify lower bounds for real-valued
  multivariate functions, defined by semialgebraic or transcendental
  expressions.  The certificate must be, eventually, formally provable
  in a proof system such as Coq.  The application range for such a
  tool is widespread; for instance Hales' proof of Kepler's conjecture
  yields thousands of inequalities. We introduce an approximation
  algorithm, which combines ideas of the max-plus basis method (in
  optimal control) and of the linear templates method developed by
  Manna et al. (in static analysis). This algorithm consists in
  bounding some of the constituents of the function by suprema of
  quadratic forms with a well chosen curvature. This leads to
  semialgebraic optimization problems, solved by sum-of-squares
  relaxations.  Templates limit the blow up of these relaxations at
  the price of coarsening the approximation.  We illustrate the
  efficiency of our framework with various examples from the
  literature and discuss the interfacing with Coq.
\end{abstract}

%\textbf{Motivation}: The aim of this work is to compare different approaches when computing lower bounds of multivariate real functions. In Section~\ref{sec:samp}, we explain how to get coarse lower bounds of multivariate polynomials when their number of variables is large. The Max-plus approach is combined together with fast minimal eigenvalues computations. In Section~\ref{sec:minimax}, we build tight approximations of transcendental multivariate approximations by using the Sollya software. The Flyspeck functions being defined by composition of univariate transcendental functions with semialgebraic functions, our framework is the following:  for each transcendental univariate elementary function $f \in \mathcal{T}$,  such as $\arctan$, $\exp$, \etc{}, defined on a real bounded interval $I$, we use the Remez algorithm to obtain the minimax polynomial $\hat{f}$ of $f$ on $I$ with the desired degree. If the argument $c$ of $f$ is semialgebraic, we compose $\hat{f}$ with $c$ to obtain a semialgebraic approximation of ${f \circ c}$. Otherwise, we apply the same algorithm recursively to $c$. Finally, we get a semialgebraic approximation of the initial function and we compute its lower bound using SOS solvers.


\section{INTRODUCTION}
\label{sect:intro}

Numerous problems coming from various fields boil down to the
computation of a certified lower bound for a real-valued multivariate
function \linebreak $f : \R^n \to \R$ over a compact semialgebraic set
$K \subset \R^n$.

Our aim is to automatically provide lower bounds for the following
global optimization problem:
\begin{equation}
\label{eq:f}
f^*  :=  \inf_{\xb \in K} f (\xb) \enspace,
\end{equation}
We want these bounds to be certifiable, meaning that their
correctness must be, eventually, formally provable in a proof system
such as Coq.  One among many applications is the set of several
thousands of non-linear inequalities which occur in Thomas Hales'
proof of Kepler's conjecture, which is formalized in the Flyspeck
project~\cite{DBLP:journals/dcg/HalesHMNOZ10,hales:DSP:2006:432}. Several
inequalities issued from Flyspeck actually deal with special cases
of Problem~\eqref{eq:f}. For instance, $f$ may be a multivariate
polynomial (polynomial optimization problems (POP)), or belong to the
algebra $\mathcal{A}$ of semialgebraic functions which extends
multivariate polynomials with arbitrary compositions of
$(\cdot)^{p}, (\cdot)^{\frac{1}{p}} (p \in \N_0), 
\lvert\cdot\rvert, +, -, \times, /, \sup(\cdot,\cdot),
\inf(\cdot,\cdot) $ (semialgebraic optimization problems),
or involve transcendental functions ($\sin$, $\arctan$, \etc).
 
Formal methods that produce precise bounds are mandatory because of the tightness of these inequalities. However, we also need to tackle scalability issues, which arise when one wants to provide coarser lower bounds for optimization problems with a larger number of variables or polynomial inequalities of a higher degree, etc.
%The
%scalability is strongly related to the number of polynomial
%inequalities and the variables of the system. 
A common idea to handle
Problem~\eqref{eq:f} is to first approximate $f$ by multivariate
polynomials through a semialgebraic relaxation and then obtain a lower
bound of the resulting POP with a specialized software. This implies
being able to also certify the approximation error in order to
conclude. Such techniques rely on hybrid symbolic-numeric
certification methods, see Peyrl and Parrilo~\cite{DBLP:journals/tcs/PeyrlP08} and Kaltofen et al.%
%, Peyrl and
%Parrilo
~\cite{KLYZ09}. They allow one to produce positivity {\em certificates}
for such POP which can be checked in proof assistants such as
Coq~\cite{Monniaux_Corbineau_ITP2011,Besson:2006:FRA:1789277.1789281},
HOL-light~\cite{harrison-sos} or
MetiTarski~\cite{Akbarpour:2010:MAT:1731476.1731498}. Recent efforts
have been made to perform a formal verification of several Flyspeck
inequalities with Taylor interval
approximations~\cite{DBLP:journals/corr/abs-1301-1702}. We also mention procedures that solve SMT problems over the real numbers, using interval constraint propagation\cite{DBLP:journals/corr/abs-1204-3513}.

Solving POP is already a hard problem, which has been extensively studied. Semidefinite programming (SDP) relaxations based methods
have been developed
by Lasserre~\cite{DBLP:journals/siamjo/Lasserre01}
and Parrilo~\cite{parrilo:polynomials}.
%~\cite{DBLP:journals/tcs/PeyrlP08}.
A sparse refinement of the hierarchy of SDP relaxations by
Kojima~\cite{Waki06sumsof} has been implemented in the SparsePOP
solver. Other approaches are based on Bernstein
polynomials~\cite{Zumkeller:2008:Thesis}, global optimization by
interval methods (see e.g.~\cite{DBLP:journals/rc/Hansen06}), branch
and bound methods with Taylor
models~\cite{DBLP:journals/mp/CartisGT11}.

%In what follows, we will consider the following running example taken from Hales' proof:
Inequalities involving transcendental functions are typically
difficult to solve with interval arithmetic, in particular due to the
correlation between arguments of unary functions (e.g. $\sin$)
or binary operations (e.g. $ +, -, \times, /$).
For illustration purpose, we consider the following running example
coming from the global optimization literature:
\begin{example}[Modified Schwefel Problem 43 from Appendix B in~\cite{Ali:2005:NES:1071327.1071336}]
\label{ex:sin} 
\[ \min_{\xb \in [1, 500]^n} f(\xb) = - \sum_{i = 1}^n (x_i + \epsilon x_{i + 1}) \sin (\sqrt{x_i}), \]
\end{example}
where $x_{n + 1} = x_1$, and $\epsilon$ is a fixed parameter in $\{0,1\}$. In the original problem, $\epsilon =
0$, \ie\ the objective function $f$ is the sum of independent
functions involving a single variable. This property may be exploited
by a global optimization solver by reducing it to the problem
$\min_{x \in [1, 500]} x \sin (\sqrt{x})$. Hence, we also consider a
modified version of this problem with $\epsilon = 1$.
%The second one is a lemma taken from Hales' proof (Lemma$_{9922699028}$ Flyspeck).

%%Thus, SDP relaxations are a powerful tool to get tight certified lower bound
%%for semialgebraic optimization problems. However, their applicability
%%is limited to small or medium size problems, since their execution
%%time grows exponentially with the relaxation order. Moreover, many 
%%applications require certifying inequalities involving both
%%transcendental functions and semialgebraic functions.
%%Standard approaches involve approximating
%%these functions by polynomials of a suitable degree,
%%which may also increase the needed relaxation order. \textbf{[XA: le paragraphe est un peu décorellé du reste\dots]}
%%
\subsubsection*{Contributions.} 
In this paper, we present an exact certification method,
aiming at handling the approximation of transcendental functions
and increasing the size of certifiable instances.
%which targets larger scale instances, 
% which extends the previous method
%of~\cite{victorecc}. semialgebraic
It consists in combining SDP relaxations \`a la Lasserre / Parrilo, with
an abstraction or approximation method. The latter is inspired by
the linear template method of Sankaranarayanan, Sipma and Manna
in static analysis~\cite{Sankaranarayana+others/05/Scalable},
its nonlinear extension by Adj\'e et al.~\cite{adjegaubertgoubault11},
and the maxplus basis method in optimal control
introduced by Fleming and McEneaney~\cite{a5}, and
developed by several authors~\cite{a6,a7,curseofdim,conf/cdc/GaubertMQ11}.
%PhysRevA.82.042319,% ref suppressed to save space
%
%of semialgebraic problems with non-linear templates based on max-plus
%underestimators. The efficiency and the scalability of the method
%hinge upon an adaptive approximation of some constituents of $f$ by
%quadratic functions.

The non-linear template method is a refinement of polyhedral based
methods in static analysis. It allows one
to determine invariants of programs
by considering a parametric family of sets, 
$S(\alpha)=\{x\mid w_i(x)\leq \alpha_i, 1\leq i\leq p\}$, 
where the vector $\alpha\in \mathbb{R}^p$ is the parameter,
and $w_1,\dots,w_p$ (the template) are fixed 
possibly non-linear functions, tailored 
to the program characteristics. The max-plus basis
method is equivalent to the approximation of the epigraph
of a function by a set $S(\alpha)$. In most basic examples, the functions
$w_i$ of the template are linear or quadratic functions. 

In the present application, templates are used
both to approximate transcendental functions,
and to produce coarser but still tractable relaxations
when the standard SDP relaxation of the semialgebraic
problem is too complex to be handled. Indeed,
SDP relaxations are a powerful tool to get tight certified lower bound
for semialgebraic optimization problems, but their applicability
is so far limited to small or medium size problems: their execution
time grows exponentially with the relaxation order, which itself
grows with the degree of the polynomials to be handled. Templates
allow one to reduce these degrees, by approximating certain projections
of the feasible set by a moderate number of nonconvex quadratic inequalities.

Note that by taking a trivial template (bound constraints, \ie{}, 
functions of the form $w_i(x)=\pm x_i$),
the template method specializes to a version of interval calculus, in which
bounds are derived by SDP techniques. By comparison, templates allow one
to get tighter bounds, taking into account the correlations
between the different variables. They are also useful as a
replacement of standard Taylor approximations of transcendental
functions: instead of increasing the degree of the approximation,
one increases the number of functions in the template.
A geometrical way to interpret the method is to think of it
in terms of ``quadratic cuts'': quadratic inequalities
are successively added to approximate the graph
of a transcendental function.

\if{
The idea of max-plus approximation comes
from optimal control: it was originally introduced
by Fleming and McEneaney~\cite{a5}, and developed
by several authors~\cite{a6,a7,curseofdim,PhysRevA.82.042319,conf/cdc/GaubertMQ11}, to represent the value function
by a ``max-plus linear combination'', which
is a supremum of certain basis
functions, like quadratic forms. 
When applied to the present context, this idea leads to
approximate from above and from below
every transcendental function appearing
in the description of the problem
by infima and suprema of finitely
many quadratic forms. 
%By Legendre-Fenchel duality,
%under semiconvexity and semiconcavity assumptions,
%this leads to a convergent approximation scheme.
In that way, we are reduced to a converging sequence
of semialgebraic problems. A geometrical way
to interpret the method is to think of it
in terms of ``quadratic cuts'': quadratic inequalities
are successively added to approximate the graph
of a transcendental function.

% The paper is structured as follows. In Section~\ref{sect:pre} we
% introduce notations and preliminaries about Semidefinite Programming
% (SDP) relaxations to explain how to deal with constrained
% polynomials optimization.  In Section~\ref{sect:tr}, we build a
% hierarchy of semialgebraic relaxations to approximate transcendental
% functions. Then, the feasible points generated by these methods are
% used to refine iteratively the semialgebraic approximations of
% transcendental functions until the needed accuracy is
% reached. Inequalities are checked by global optimization methods
% that is, solving a hierarchy of relaxed problems using Sum of Square
% (SOS) solvers such as SparsePOP. This solver can be interfaced with
% several SDP solvers (e.g. SeDuMi~\cite{Sturm98usingsedumi},
% CSDP~\cite{Borchers97csdp}, SDPA~\cite{YaFuNaNaFuKoGo:10}) to solve
% a hierarchy of SDP problems by Lasserre with a sparse formulation.

%We advance the following new lines of research:
%\begin{itemize}	
%\item 

The proposed method (Figure~\ref{alg:algo_{template_optim}}) may be
summarized as follows.  Let $f$ be a function and $K$ a box issued
from a Flyspeck inequality, where $f$ belongs to the set of
transcendental functions obtained by composition of semialgebraic
functions with $\arctan$, $\arccos$, $\arcsin$, $\exp$, $\log$,
$|\cdot|$, $(\cdot)^{\frac{1}{p}} (p \in \N_0) $, $ +, -, \times, /,
\sup(\cdot,\cdot), \inf(\cdot,\cdot)$. We alternate steps of
approximation in which an additional quadratic function is added to
the representation, and optimization steps in which an SDP relaxation
from Lasserre hierarchy is solved. The information on the location of
the optimum inferred from this relaxation is then used to refine
dynamically the quadratic approximation.  

\if{
To solve the resulting Problem~\eqref{eq:f},
  Section~\ref{sect:tr} describes how to underestimate $f$ by a
  semialgebraic function $f_{sa}$ on a compact set $\Ksa$ that
  contains $K$. Then, the feasible points generated by these methods
  are used to refine iteratively the semialgebraic approximations of
  transcendental functions until the needed accuracy is reached. Hence
  it leads to the resolution of a hierarchy of
  Problem~\eqref{eq:f_{sa}} instances. We define this hierarchy by
  adding quadratic constraints to a semialgebraic optimization
  problem. In other words, we optimize an objective function on a
  sequence of semialgebraic relaxations obtained by applying
  successive quadratic cuts on the initial set $K$.
%\item  To the best of our knowledge, there is no available software implementation that solves the general Problem~\eqref{eq:f_{sa}} using SDP relaxations.
%\item
  Based on the properties of these
  relaxations, the following inequality holds:}\fi 

This way, at each step of the algorithm, we refine the following
inequalities
	\begin{equation}
	\label{eq:3relax}
		f^* \geq f_{sa}^* \geq f_{pop}^* \enspace,
	\end{equation}
where $f^*$ is the optimal value of the original problem,
$f^*_{sa}$ the optimal value of its current semialgebraic approximation
and $f_{pop}^*$ the optimal value of the SDP relaxation
which we solve. The lower estimate $f_{pop}^*$ does
converge to $f^*$. This follows from a theorem
of Lasserre (convergence of moment SDP relaxations)
and from the consistency of max-plus approximation, see Theorem~\ref{th:transc}.}\fi

%%An important issue, for the practical efficiency of the method, is the
%%simultaneous tuning of the precision of the max-plus approximation and
%%of the orders of SDP relaxations. In the present method,
%%
%%Here, we compute lower bounds of $f$
%%by solving a semialgebraic optimization problem on a domain which is
%%defined by sub-level sets of quadratic forms. Our method generalizes
%%the linear templates method by Manna, Sankaranarayanan and
%%Spima~\cite{Sankaranarayana+others/05/Scalable}. It allows to reduce
%%the number of lifting variables by adding quadratic cuts to the
%%semialgebraic optimization problems that we solve. If the number of
%%lifting variables or the number of quadratic cuts becomes too large we
%%define new max-plus approximations for some constituents of $f$. A
%%first collection $\mathcal{C}_1$ of max-plus approximations is built
%%in at each univariate transcendental node. If $\mathcal{C}_1$ involves
%%too many variables or quadratic cuts, we define a second collection
%%$\mathcal{C}_2$ of approximations. We determine this new collection
%%$\mathcal{C}_2$ by solving semialgebraic optimization problems. The
%%feasible set of these problems come from the polynomial inequalities
%%related to $\mathcal{C}_1$. Afterwards, we use $\mathcal{C}_2$ to
%%approximate some constituents of $f$ and solve optimization problems
%%involving less variables and/or less quadratic cuts (see
%%Figure~\ref{alg:samp_template}). %Hence, we ensure the scalability of the method.
%%
%%TODO: reformuler ou enlever partie sur tuning des approx max-plus
\if{
How to perform optimally this tuning and the choice of the templates is still not well understood.
However, we present experimental results, both
for some elementary examples as well as non-linear inequalities issued from the Flyspeck project, giving some indication that certain hard subclasses of problems (sum of arctan of correlated functions in many variables) can be solved in a scalable way.
}\fi

\if{
Max-plus approximation has attracted interest because it may attenuate the ``curse of dimensionality'' for some structured
problems~\cite{mccomplex}.  Indeed, the estimate
of~\cite{conf/cdc/GaubertMQ11} shows that the number of quadratic terms needed to reach an $\epsilon$-approximation of a function of $d$ variables is of order $\epsilon^{-d/2}$, where $d$ is the dimension.
Hence, max-plus approximations can be applied to fixed, small
dimensional sub-expressions of complex high dimensional expressions, in a curse of dimensionality free way. In particular, in the Flyspeck inequalities involve generally 6 variables, but only univariate transcendental functions, so $d=1$.

An alternative, more standard approach, would be to approximate
transcendental functions by polynomials of a sufficiently high
degree, and to apply SDP relaxations to the polynomial problems
obtained in this way. Some preliminary experiments indicated
that this method is not scalable. Another alternative approach, which is quite effective on Flyspeck type inequalities,
is to run branch and bound type algorithms with interval arithmetic. However, in some instances, this leads to certifying an exponential number of interval arithmetic computations. Thus, it is of interest to investigate hybrid methods such as the present one, in order to obtain more concise certificates.
}\fi

\if{
To solve the POP instances, several solvers are available as Gloptipoly~\cite{henrion:hal-00172442} or 
Kojima sparse refinement of the hierarchy of SDP relaxations~\cite{Waki06sumsof},
implemented in the SparsePOP solver~\cite{DBLP:journals/toms/WakiKKMS08}.
These solvers are interfaced with several SDP solvers
  (e.g. SeDuMi~\cite{Sturm98usingsedumi}, CSDP~\cite{Borchers97csdp},
  SDPA~\cite{YaFuNaNaFuKoGo:10}).  
}\fi

The present paper is a followup of~\cite{victorecc}, in which
the idea of  max-plus approximation of transcendental function was applied to formal proof. By comparison, the new ingredient is the introduction of the template technique (approximating projections of the feasible sets), leading to
an increase in scalability.

The paper is organized as follows. In Section~\ref{sect:pre}, we
recall the definition and properties of Lasserre relaxations of
polynomial problems (Section~\ref{subsect:pop}), together with reformulations by Lasserre and
Putinar of semialgebraic problems classes. In
Section~\ref{subsect:hybrid}, we outline the conversion of the numerical
SOS produced by the SDP solvers into an exact rational
certificate. Then we explain how to verify this certificate in Coq. The max-plus approximation, and the main algorithm based on the non-linear templates method are presented in Section~\ref{sect:template}. 
Numerical results are presented in Section~\ref{sect:bench}. We demonstrate the scalability of our approach by certifying bounds of 
non-linear problems involving up to $10^3$ variables, as well as
non trivial inequalities issued from the Flyspeck project.

\section{NOTATION AND PRELIMINARY RESULTS}\label{sect:pre}
\if{
We consider the vector space $\mathcal{S}_{n}$ of symmetric $n \times
n$ matrices. 
%It is equipped with the usual inner product $\langle X, \, Y \rangle = \text{Tr} (XY)$ for $X$, $Y \in \mathcal{S}_{n}$. The Frobenius norm of a matrix $X \in \mathcal{S}_{n}$ is defined by $\Vert X \Vert_{F} := \sqrt{\text{Tr} (X^2)}$. 
A matrix $M \in \mathcal{S}_{n}$ is called positive semidefinite if $x^T M x \geq 0, \, \forall x \in \R^n$. In this case, we write $M \succcurlyeq 0$, and
define a partial order by writing $X \succcurlyeq Y$ (resp.\ $X \succ Y$) if and only if $X - Y$ is positive semidefinite (resp.\ positive definite).
}\fi
%\hspace{5 mm}   
\if{

Let $\mathcal{B}_d$ denote the basis of monomials for the $d$-degree
real-valued polynomials in $n$ variables :
%	\begin{equation}
%	\label{eq:monbasis}
%		1, x_1, x_2, \dots , x_1^2, x_1 x_2, \dots , x_1 x_n, x_2 x_3, \dots, x_n^2, \dots , x_1^d, \dots, x_n^d
%	\end{equation}
\begin{equation}
	\label{eq:monbasis}
		1, x_1, x_2, \dots , x_1^2, x_1 x_2, \dots, x_n^2, \dots, x_n^d
	\end{equation}
}\fi
%A symmetric matrix $W$ is called positive semidefinite if all its eigenvalues are non-negative. In this case, we write $W \succcurlyeq 0$. 
Let $\R_d[\xb]$ be the vector space of multivariate polynomials in $n$ variables of degree $d$ and $\R[\xb]$ the set of multivariate
polynomials in $n$ variables.  We also define the cone of sums of squares of degree at most $2 d$:
	\begin{equation}
%	\label{eq:cone_sos}
%	\begin{align}
	\Sigma_{d} [\xb] = \Bigl\{\,\sum_i q_i^2, \, \text{ with } q_i \in \R_d[\xb] \,\Bigr\}.
%	\end{align}  
	\end{equation}
        The set $\Sigma_{d} [\xb]$ is a closed, fully dimensional
        convex cone in $\R_{2 d}[\xb]$. We denote by $\Sigma[\xb]$ the
        cone of sums of squares of polynomials in $n$ variables.  

        \if{ If $\fpop : \R^n \to \R$ is a $d$-degree multivariate
          polynomial, we write
	\begin{equation}
	\label{eq:pdecomp}
	\fpop(\xb) = \sum_{\alpha} p_{\alpha}x^{\alpha},  
	\end{equation}
	$\text{with} \quad x^{\alpha} := x_1^{\alpha_1} \dots x_n^{\alpha_n} \quad \text{and} \quad \alpha \in \mathcal{F}_d := \{ \alpha \in \N^n : \quad  \sum_{i} \alpha_i \leq d \}$
}\fi

\subsection{Constrained Polynomial Optimization Problems and SDP}
\label{subsect:pop}\label{sect:sdp}
We consider the general constrained polynomial optimization problem (POP):
\begin{equation}
\label{eq:cons_pop}
\fpop^*  :=  \inf_{\xb \in \Kpop} \fpop (\xb),
\end{equation}
where $\fpop : \R^n \to \R$ is a $d$-degree multivariate polynomial,
$\Kpop$ is a compact set defined by inequalities $g_1(\xb)
\geq 0,\dots,g_m(\xb) \geq 0$, where $g_j(\xb) : \R^n \to \R$ is a
real-valued polynomial of degree $\omega_j$, for $j = 1,\dots,m$.
Recall that the {\em set of feasible points} of an optimization
problem is simply the domain over which the optimum is taken,
i.e., here, $\Kpop$.
% the {\em feasible set} of Problem~\eqref{eq:cons_pop}.

\subsubsection{Lasserre's hierarchy of semidefinite relaxations.}
We set $g_0 := 1$ and take
 $k \geq k_0 := \max(\lceil d / 2 \rceil ,\max_{1 \leq j \leq m} 
\lceil \omega_j / 2\rceil)$. We consider the following
hierarchy of semidefinite relaxations for Problem~\eqref{eq:cons_pop},
consisting of the optimization problems $Q_k$, $k\geq k_0$,
\[
			Q_k:\left\{			
			\begin{array}{ll}
			 \sup_{\mu, \sigma_j} & \mu \\			 
			 \text{s.t.} & \fpop (\xb) - \mu = \sum_{j = 0}^m \sigma_j(\xb) g_j(\xb), \\
			  & \mu\in \mathbb{R},\qquad \sigma_j \in \Sigma_{k - \lceil \omega_j / 2 \rceil} [\xb], j = 0,\cdots,m .\\
			\end{array} \right.
			\]
We denote by $\sup (Q_k)$ the optimal value of $Q_k$. A feasible point $(\mu,\sigma_0,\dots,\sigma_m)$ of Problem $Q_k$ is said to be a {\em SOS certificate},
showing the implication $g_1(\xb)\geq 0,\dots,g_m(\xb) \geq 0 
\implies \fpop(\xb)\geq \mu$. 

The sequence of optimal values $(\sup (Q_k))_{k \geq k_0}$ is non-decreasing.
Lasserre showed~\cite{DBLP:journals/siamjo/Lasserre01} 
that it does converge to $\fpop^*$ under certain assumptions
on the polynomials $g_j$. Here, 
%that  Under a certain assumption, it Here, we 
we will consider sets $\Kpop$ included in a box of $\mathbb{R}^n$, so that Lasserre's assumptions are automatically satisfied.

%This assumption holds in our case since the non-linear global optimization problems that we consider .
%We refer the reader to~\cite{DBLP:journals/siamjo/Schweighofer05} for a comprehensive discussion about equivalent statements for this assumption.  One of these conditions holds in our case since the non-linear inequalities to be proved in the Flyspeck project typically involve a variable $x$ lying in a box $K \subset \R^n$. It holds also for the problems of the global optimization literature that we consider.

\if{
\begin{example}[from $\text{Lemma}_{4717061266}$ Flyspeck]
\label{ex:4717}
Let $\Delta x := x_1 x_4 ( - x_1 +  x_2 +  x_3  - x_4 +  x_5 +  x_6) 
+ x_2 x_5 (x_1  -  x_2 +  x_3 +  x_4  - x_5 +  x_6) 
+ x_3 x_6 (x_1 +  x_2  -  x_3 +  x_4 +  x_5  -  x_6)
- x_2 x_3 x_4  -  x_1 x_3 x_5  -  x_1 x_2 x_6  - x_4 x_5 x_6$. Using SparsePOP~\cite{DBLP:journals/toms/WakiKKMS08},
the optimal value of $128$ for
the problem $ \inf_{x \in [4, 6.3504]^6} \Delta x$ is obtained at
the $Q_2$ relaxation with $\epsilon_{SDP} = 10^{-8}$.
\end{example}
%\begin{proof}
%\end{proof}
}\fi
\subsubsection{Application to semialgebraic optimization.}
\label{sect:sa}

Given a semialgebraic function $\fsa$, we consider the problem $\fsa^* = \inf_{\xb \in \Ksa} \fsa (\xb)$, where $\Ksa$ is a basic semialgebraic set. Moreover, we assume that $\fsa$ has a basic semialgebraic lifting (for more
details, see e.g.~\cite{DBLP:journals/siamjo/LasserreP10}). This implies that
we can add auxiliary variables $z_1,\dots,z_p$ (lifting variables),
%there exists $p, s \in \N$, 
and construct polynomials $ h_1, \dots , h_s \in \R[\xb,
z_1,\dots,z_p]$ defining the semialgebraic set $\Kpop := \{ (\xb, \, z_1,\dots,z_p) \in \R^{n+p} : \xb \in \Ksa, 
 h_1(\xb, \zb) \geq 0,\dots, h_s(\xb, \zb) \geq 0 \}$, such that $\fpop^* := \inf_{(\xb, \zb) \in \Kpop} z_p$ is a lower bound of $\fsa^*$.
 $\fsa^* := \inf_{(\xb, \zb) \in \Kpop} z_p$
 \if{
\begin{multline*}
\Kpop :=  \{ (\xb, \, z_1,\dots,z_p) \in \R^{n+p} : \xb \in \Ksa, 
 h_1(\xb, \zb) \geq 0,\dots, h_s(\xb, \zb) \geq 0 \}
\end{multline*}
}\fi

\if{
\begin{example} [from Lemma$_{9922699028}$ Flyspeck]
\label{ex:atn1}
Let $\Delta \xb := x_1 x_4 ( - x_1 +  x_2 +  x_3  - x_4 +  x_5 +  x_6) 
+ x_2 x_5 (x_1  -  x_2 +  x_3 +  x_4  - x_5 +  x_6) 
+ x_3 x_6 (x_1 +  x_2  -  x_3 +  x_4 +  x_5  -  x_6)
- x_2 x_3 x_4  -  x_1 x_3 x_5  -  x_1 x_2 x_6  - x_4 x_5 x_6$. We
consider the function $\fsa := \frac{\partial_4 \Delta \xb }{\sqrt{4 x_1
    \Delta \xb}}$ and the set $\Ksa := [4, 6.3504]^3 \times [6.3504, 8]
\times [4, 6.3504]^2$. We introduce two lifting variables $z_1$ and
$z_2$, respectively representing the terms $\sqrt{4 x_1 \Delta \xb}$ and
$\frac{\partial_4 \Delta \xb }{\sqrt{4 x_1 \Delta \xb}}$.
% We also use a lower bound $m_1$ of $ \inf_{x \in \Ksa} \sqrt{4 x_1
% \Delta x}$ and an upper bound $M_1$ of $ \sup_{x \in \Ksa} \sqrt{4
% x_1 \Delta x}$ which can be both computed by solving auxiliary
% subproblems.
We add bound constraints over $z_1$ by solving auxiliary subproblems.
We define $\Kpop$ by:
\begin{align*}
\Kpop := {} & \{ (\xb, z_1 ,z_2) \in \R^{6+2} : 
\xb \in \Ksa, h_j(\xb, z_1, z_2) \geq 0, j = 1,\dots,7 \},
\end{align*}
where the multivariate polynomials $h_j$ are defined by:
\begin{align*}
h_1 & := z_1 - m_1 & h_2 & := M_1 - z_1 & h_3 & := z_1^2 - 4 x_1 \Delta x & h_4 & := - z_1^2 + 4 x_1 \Delta x \\
h_5 & := z_1 & h_6 & := z_2 z_1 - \partial_4 \Delta x & h_7 & := - z_2 z_1 + \partial_4 \Delta x
\end{align*}

%%We obtain a hierarchy of semidefinite relaxations $Q_k^{sa}$ for the
%%the problem $\fsa^* = \inf_{\xb \in \Ksa} \fsa (\xb)$, 
%%by specializing the relaxations $Q_k$
%%of Section~\ref{subsect:pop} to the latter sets $\Kpop$.

 The lower bound $m_2 = -0.618$ computed by the
%^{sa}
relaxation $Q_2$ is too coarse. A tighter lower bound $m_3 = -0.445$ is
obtained by the third relaxation, but it requires ten times more CPU time.
%, but it consumes more CPU time.
\end{example} 
}\fi
%\subsection{Complexity Considerations}
%\label{sect:decrease}
%The size of the truncated moment SDP matrices grows exponentially with
%the SDP-relaxation order and polynomially with the number of
%variables. Indeed, for a $n$ variables polynomial optimization
%problem, the relaxation $Q_k$ involves $O (n^ {2 k})$ SDP
%moment variables and linear matrix inequalities (LMIs) of size
%$O (n^ k)$.

\if{ There are several ways to decrease the size of these
  matrices. First, symmetries in SDP relaxations for polynomial
  optimization problems can be exploited to replace one SDP problem
  $Q_k$ of size $O (n^ {2 k})$ by several smaller
  SDPs~\cite{DBLP:journals/corr/abs-1103-0486}. Notice that it is
  possible only if the multivariate polynomials of the initial problem
  are invariant under the action of a finite subgroup $G$ of the group
  $GL_n(\R)$. Furthermore, one can exploit the structured sparsity of
  the problem to replace one SDP problem $Q_k$ of size $O (n^ {2 k})$
  by an SDP problem of size $O (\kappa^ {2 k})$ where $\kappa$ is the
  average size of the maximal cliques correlation pattern of the
  polynomial variables (see~\cite{DBLP:journals/toms/WakiKKMS08}).
}\fi

\subsection{Hybrid Symbolic-Numeric Certification and Formalization}
\label{subsect:hybrid}

The previous relaxation $Q_k$ can be solved with several semidefinite
programming solvers (e.g. SDPA~\cite{YaFuNaNaFuKoGo:10}). These
solvers are implemented using floating-point arithmetics. In order to
build formal proofs, we currently rely on  exact
rational certificates which are needed to make formal proofs:
Coq, being built on a computational formalism, is well equipped for
checking the correctness of such certificates.

Such rational certificates can be obtained by a rounding and
projection algorithm of Peyrl and
Parillo~\cite{DBLP:journals/tcs/PeyrlP08}, with an improvement of
Kaltofen et al.~\cite{KLYZ09}.  Note that if the SDP formulation of
$Q_k$ is not strictly feasible, then the rounding and projection
algorithm fails. However, Monniaux and Corbineau proposed a partial
workaround for this issue~\cite{Monniaux_Corbineau_ITP2011}.  In this
way, except in degenerate situations, we arrive at a candidate SOS
certificate with rational coefficients,
$(\mu,\sigma_0,\dots,\sigma_m)$.  This certificate can 
straightforwardly be translated to Coq; the verification then boils
down to formally checking that this SOS certificate does satisfy the
equality constraint in $Q_k$ with Coq's $\mathtt{field}$ tactic, which implies that $\fpop^* \geq {\mu}$.
This checking is typically handled by generating Coq scripts from the OCaml framework, when the lower bound $\mu$ obtained at the relaxation $Q_k$ is accurate enough.

Future improvements could build, for instance, on future Coq libraries
handling algebraic numbers or future tools to better handle floating
point approximations inside Coq.

\if{on the hyperplane $\chi := \{\sigma' \mid \fpop (\xb) - \mu^* \simeq \sum_{j = 0}^m \sigma'_j(\xb) g_j(\xb)
, \sigma'_j \in \Sigma_{2 k - w_j} [x], j = 0,\cdots,m. \}$ 
}\fi

\section{MAX-PLUS APPROXIMATIONS AND NON-LINEAR TEMPLATES}
\label{sect:template}

%In this section, we introduce an algorithm to produce a hierarchy of
%lower bounds for global optimization problems
%(Problem~\eqref{eq:f}). The algorithm relies on an adaptive
%basic-semialgebraic relaxation, in which approximations of some
%constituents of $f$ are iteratively refined.
	
\subsection{Max-plus Approximations and Non-linear Templates}
\label{subsect:maxplus}
%TODO: expliquer non-linear templates method
The max-plus basis method in optimal control~\cite{a5,a6,curseofdim}
involves the approximation from below of a function $f$ in $n$ variables
by a supremum
\begin{align}
f \gtrapprox g:=\sup_{1\leq i \leq p} \lambda_i + w_i  \enspace .\label{e-sup}
\end{align}
The functions $w_i$ are fixed in advance, or dynamically
adapted by exploiting the problem structure. The parameters
$\lambda_i$ are degrees of freedom. 
%(a function of a possibly large number $n$ of variables
%is coarsely approximated by a function with a fixed number $p$ of parameters).

This method is closely related to the non-linear extension~\cite{adjegaubertgoubault11} of the template method~\cite{Sankaranarayana+others/05/Scalable}. This
extension deals with parametric families of subsets of $\mathbb{R}^n$ of the form
\(
S(\alpha) = \{x\mid w_i(x) \leq \alpha_i, \; 1\leq i\leq p\}.
\)
The template method consists in propagating approximations of the
set of reachables values of the variables of a program
by sets of the form $S(\alpha)$.  
The non-linear template and max-plus approximation methods
are somehow equivalent. Indeed,
the $0$-level set of $g$, $\{x\mid g(x) \leq 0\}$, is nothing but
$S(-\lambda)$, so templates can be recovered
from max-plus approximations, and vice versa.

The functions $w_i$ are usually required to be 
quadratic forms,
\[
w_i(x)  = p_i^\top x + \frac{1}{2} x^\top A_i x \enspace,
\]
where $p_i\in \mathbb{R}^n$ and $A_i$ is a symmetric matrix. A
basic choice is  $A_i=-c I$, where $c$ is a fixed constant, and $I$ the identity matrix. Then, the parameters $p$ remain the only degrees of freedom.

The consistency of the approximation
follows from results of Legendre-Fenchel duality. Recall
that a function $f$ is said to be {\em $c$-semiconvex} if $x\mapsto f(x)+ c\|x\|^2$ is convex. Then, if $f$ is $c$-semiconvex and lowersemicontinuous,
as the number of basis functions $r$ grows, the best approximation $g\lessapprox f$ 
by a supremum of functions of type~\eqref{e-sup}, with $A_i=-cI$,
is known to converge to $f$~\cite{a5}.
The same is true without semiconvexity assumptions
if one allows $A_i$ to vary~\cite{agk04short}. 

%%, which imply that an arbitrary lsc function $f$
%%has an infinite max-plus representation by quadratic forms, and that
%%the curvature of these quadratic forms can be bounded when $f$ is known
%%to be $c$-semiconvex, meaning that $c\|x\|^2/2 + f$ is convex. The
%%second part of the next theorem is a standard result on
%%Legendre-Fenchel duality, the first one is proved in~\cite{}.
%%\begin{theorem}[Consistency theorem, Corollary of Legendre-Fenchel duality, see~\cite{}]
%%Any lsc function can be written as a supremum of quadratic forms.
%%Any $c$-semiconvex lsc functions can be written as a supremum
%%of quadratic forms with Hessian matrix $-c I$.
%%\end{theorem}
A basic question is to estimate the number of basis functions needed to
attain a prescribed accuracy. A typical
result is proved in~\cite[Theorem~3.2]{conf/cdc/GaubertMQ11},
as a corollary of techniques of Gr\"uber concerning the approximation
of convex bodies by circumscribed polytopes. 
This theorem shows that if $f$ is $c-\epsilon$ semiconvex,
for $\epsilon>0$, twice continuously differentiable,
and if $X$ is a full dimensional compact
convex subset of $\mathbb{R}^n$, then, the best approximation
$g$ of $f$ as a supremum or $r$ functions as in~\eqref{e-sup}, with 
$w_i(x) = p_i^\top x-c\|x\|^2/2$, satisfies
\begin{align}
\|f-g\|_{L_\infty(X)} \simeq \frac{C(f)}{r^{2/n}}
\label{e-pessimistic}
\end{align}
where the constant $C(f)$ is explicit (it depends of $\det (f''+ cI)$
and is bounded away from $0$ when $\epsilon$ is fixed). This estimate
indicates that some curse of dimensionality is unavoidable:
to get a uniform error of order $\epsilon$, one needs
a number of basis functions of order $1/\epsilon^{n/2}$. However,
in what follows, we shall always apply the approximation 
to small dimensional constituents of the optimization
problems ($n=1$ when one needs to approximate transcendental
functions in a single variable). We shall also apply the
approximation by templates to certain relevant
small dimensional projections of the set of lifted variables,
leading to a smaller effective $n$.
Note also that for optimization purposes, a uniform approximation
is not needed (one only needs an approximation tight enough
near the optimum, for which fewer basis
functions are enough).

\if{
Let $\cB$ be a set of functions $\R^d\to \R$, whose elements
will be called \textit{max-plus basis functions}. Given
a function $f:\R^d\to \R$, we look
for a representation of $f$ as a linear combination
of basis functions in the max-plus sense, \ie{},
\begin{equation}
f = \sup_{w\in \cB}(a(w) + w) \label{e-Fenchel}
\end{equation}
where $(a(w))_{w\in \cB}$ is a family of elements of $\R
\cup\{-\infty\}$ (the ``coefficients'').  The correspondence between
the function $x \mapsto f(\xb)$ and the coefficient function $w\mapsto
a(w)$ is a well studied problem, which has appeared in various guises
(Moreau conjugacies, generalized Fenchel transforms, Galois
correspondences, see~\cite{agk04} for more background).

The idea of max-plus approximation~\cite{a5,mceneaney-livre,a6} is to
choose a space of functions $f$ and a corresponding set $\cB$ of basis
functions $w$, and to approximate from below a given $f$ in this space
by a finite max-plus linear combination, \(%\begin{align}
f \simeq \sup_{w\in \cF}(a(w) + w)
%\label{e-Fenchelapprox}
\)
%\end{align}
where $\cF\subset\cB$ is a finite subset. Note that 
$ \sup_{w\in \cF}(a(w) + w) $ is not only an approximation but a valid
lower bound of $f$. 

Following~\cite{a5,a6}, for each constant
$\gamma\in \R$, we shall consider 
the family of quadratic functions $\cB=\{w_{y}\mid y\in \R^d\}$ where 
\[
w_{y}(\xb):= -\frac{\gamma}{2}\|x-y\|^2  \enspace .
\]
Recall that a function is $\gamma$-semiconvex if and only if the
function $x\mapsto \phi(\xb)+\frac{\gamma}{2}|x|^2$ is convex. Then, it
follows from Legendre-Fenchel duality that the space of functions $f$
which can be written as~\eqref{e-Fenchel} is precisely the set of
lower semicontinuous $\gamma$-semiconvex functions.

The transcendental functions
which we consider here are twice
continuously differentiable. Hence, their
restriction to any bounded convex set 
is $\gamma$-semiconvex for a sufficiently large $\gamma$,
so that they can be approximated by finite suprema
of the form $\sup_{w\in \cF}(a(w) + w) $ with $\cF\subset \cB$.
A result of~\cite{conf/cdc/GaubertMQ11} shows that if $N=|\cF|$
basis functions are used, then the best approximation
error is $O(1/N^{2/d})$ (the error is the sup-norm, over any compact set).
For the applications considered in this paper, $d=1$.

In this way, starting from a transcendental univariate
elementary function $f \in \mathcal{T}$,  such as $\arctan$, $\exp$, \etc{},
defined on a real bounded interval $I$, we arrive at a 
semialgebraic lower bound of $f$, which is nothing but
a supremum of a finite number of quadratic functions.

\begin{example}
Consider the function $f= \arctan$ on an interval $I := [m, M]$.
For every point $a\in I$, we can find a constant $\gamma$
such that
\[
\arctan (x)  \geq \parab_{a}^-(x):=  -\frac{\gamma}{2} (x-a)^2 +f'(a) (x - a) + f (a)
\enspace .
\]
Choosing $\gamma=\sup_{x\in I} -f''(x)$ always work. However, 
it is convenient to allow $\gamma$ to depend on the choice of $a$ to get tighter lower bounds. 
Choosing a finite subset $A\subset I$, we arrive at an approximation
\begin{equation}
\label{eq:max_par}
\forall x \in I , \, \arctan \, (a) \geq \max_{a\in A} \, \parab_{a}^-(x) \enspace .
\end{equation} 

Semialgebraic overestimators $x \mapsto \min_{a \in
  A} \parab_{a}^+(x)$ can be defined in a similar way. Examples of
such underestimators and overestimators are depicted in
Fig.~\ref{fig:atn}.
\end{example}

\begin{figure}[ht]	
\begin{center}
\begin{tikzpicture}[scale=1]
\draw[->] (-3.5,0) -- (3.5,0) node[right] {$a$};
\draw[->] (0,-1.5) -- (0,2.2) node[above] {$y$};

\draw[color=darkgreen, dotted, thick] plot [domain=-2.5:1.2] (\x,
{0.32*(\x +1)^2 + 1/2* (\x +1) - 0.79 }) node[anchor = north west]
{$\parab_{a_1}^+$}; \draw[color=darkgreen, dotted, thick] plot
[domain=0.5:3] (\x, {0.32*(\x -2)^2 + 1/5* (\x -2) +1.1 }) node[anchor
= south west] {$\parab_{a_2}^+$};

\draw[color=whitegreen, dotted, thick] plot [domain=-0.3:3] (\x,
{-0.32*(\x -2)^2 + 1/5* (\x -2) + 1.1 }) node[anchor = north east]
{$\parab_{a_2}^-$}; \draw[color=whitegreen, dotted, thick] plot
[domain=-1.8:0.5] (\x, {-0.32*(\x +1)^2 + 1/2* (\x +1) -0.79 })
node[anchor = north west] {$\parab_{a_1}^-$};

\draw (2,-0.3) node {$a_2$}; \draw (2,3pt) -- (2,-3pt); \draw [black,
dashed] (2,0) -- (2,1.1071); \draw (-1.,-0.3) node {$a_1$}; \draw
(-1,3pt) -- (-1,-3pt); \draw [black, dashed] (-1,0) -- (-1,-0.7853);

\draw[color=red!80] plot [domain=-3:3] (\x,{rad(atan(\x))})
node[right] {$\arctan$}; \draw (-3,-0.3) node {$m$}; \draw (3.,-0.3)
node {$M$}; \draw (3,3pt) -- (3,-3pt); \draw (-3,3pt) -- (-3,-3pt);
\end{tikzpicture}
\caption{Semialgebraic underestimators and overestimators for $\arctan$}	\label{fig:atn}		
\end{center}
\end{figure}
}\fi

\subsection{A Templates Method based on Max-plus Approximations}
\label{subsect:alg1}

We now consider an instance of Problem~\eqref{eq:f}. We assume that
$K$ is a box and we identify the objective function $f$ with its
abstract syntax tree $t_f$. We suppose that the leaves of $t_f$ are
semialgebraic functions, and that the other nodes are either basic binary
operations ($+$, $\times$, $-$, $/$), or unary transcendental functions ($\sin$, \etc{}). %For the sake of the simplicity,
%we suppose that each univariate transcendental function is monotonic.

Our main algorithm $\templateoptim$ (Figure~\ref{alg:samp_template}) is based on a previous method of the authors~\cite{victorecc}, in which the objective function is bounded by means of semialgebraic functions. For the sake of completeness, we first recall the basic principles of this method.

\paragraph{Bounding the objective function by semialgebraic estimators.}

Given a function represented by an abstract tree $t$, semialgebraic lower and upper estimators $t^-$ and $t^+$ are computed by induction.
If the tree is reduced to a leaf, \ie\ $t \in \mathcal{A}$, it suffices to set $t^- = t^+ := t$. If the root of the tree corresponds to a binary operation $\mathtt{bop}$ with children $c_1$ and $c_2$, then the semialgebraic estimators $c_1^-$, $c_1^+$ and $c_2^-$, $c_2^+$ are  composed using a function $\mathtt{compose\_bop}$ to provide bounding estimators of $t$. Finally, if $t$ corresponds to the composition of a transcendental (unary) function $\phi$ with a child $c$, we first bound $c$ with semialgebraic functions $c^+$ and $c^-$. We compute a lower bound $c_m$ of $c^-$ as well as an upper bound  $c_M$ of $c^+$ to obtain an interval $I := [c_m, c_M]$ enclosing $c$. Then, we bound $\phi$ from above and below by computing parabola at given control points (function $\mathtt{build\_par}$), thanks to the semiconvexity properties of $\phi$ on the interval $I$. These parabola are composed with $c^+$ and $c^-$, thanks to a function denoted by $\mathtt{compose}$. 

These steps correspond to the part of the algorithm $\templateoptim$ from Lines~\lineref{line:begin_ecc} to~\lineref{line:end_ecc}.% in the algorithm $\mathtt{samp\_template}$. 

%\if{
%\begin{remark}
%\textbf{[TODO: Victor, peux-tu stp décrire un peu la fonction $\mathtt{build\_par}$ et expliquer pourquoi on a besoin des bornes sur son argument. Reprend la discussion de ECC, mais pas de copier/coller.}
%\end{remark}
%}\fi

\paragraph{Reducing the complexity of semialgebraic estimators using templates.}

The semialgebraic estimators previously computed are used to determine lower and upper bounds of the function associated with the tree $t$, at each step of the induction. The bounds are obtained by calling the functions $\minsa$ and $\maxsa$ respectively, which reduce the semialgebraic optimization problems to polynomial optimization problems by introducing extra lifting variables (see Section~\ref{sect:pre}).

However, the complexity of solving the POPs can grow significantly because of the number $\lift$ of lifting variables. If $k$ denotes the relaxation order, the corresponding SDP problem $Q_k$ indeed involve linear matrix inequalities of size $O((n +\lift)^k)$ over $O((n + \lift)^{2 k})$ variables.

Consequently, this is crucial to control the number of lifting variables, or equivalently, the complexity of the semialgebraic estimators. For this purpose, we introduce the function $\mathtt{build\_template}$. It allows to compute approximations of the tree $t$ by means of suprema/infima of quadratic functions, when the number of lifting variables exceeds a user-defined threshold value $\lift^{\max}$. The algorithm is depicted in Figure~\ref{alg:build_template}. Using a heuristics, it first builds candidate quadratic forms $q_j^-$ and $q_j^+$ approximating $t$ at each control point $\xb_j$ (function $\mathtt{build\_quadratic\_form}$, described below). Since each $q_j^-$ does not necessarily underestimate the function $t$, we then determine the lower bound $m_j^-$ of the semialgebraic function $t^- - q_j^-$, which ensures that $q_j^- + m_j^-$ is a quadratic lower-approximation of $t$. Similarly, the function $q_j^+ + M_j^+$ is an upper-approximation of $t$. The returned semialgebraic expressions  $\max_{1 \leq j \leq r} \{q_j^- + m_j^-\}$ and $\min_{1 \leq j \leq r} \{q_j^+ + M_j^+\}$ now generate only one lifting variable (representing $\max$ or $\min$).

Quadratic functions returned by $\mathtt{build\_quadratic\_form}(t,\xb_j)$ are of the form:
\[
q_{\xb_j,\lambda} : \xb \mapsto t(\xb_j) + \gradf{t}{\xb_j} \, (\xb - \xb_j) + \frac{1}{2}(\xb - \xb_j)^{T} \hessf{t}{\xb_j} \, (\xb - \xb_j) + \frac{1}{2} \lambda(\xb - \xb_j)^2
\]
(we assume that $t$ is twice differentiable) where $\lambda$ is computed as follows. We sample the Hessian
matrix difference $\hessf{t}{\xb} - \hessf{t}{\xb_j}$ over a finite set
of random points $R \subset K$, and construct a matrix interval $D$ enclosing all the entries of $(\hessf{t}{\xb} - \hessf{t}{\xb_j})$ for $\xb \in R$. A lower bound $\lambda^-$ of the
minimal eigenvalue of $D$ is obtained by applying a robust SDP method
on interval matrix described by Calafiore and Dabbene
in~\cite{springerlink:10.1007/s10957-008-9423-1}. Similarly, we get an
upper bound $\lambda^+$ of the maximal eigenvalue of $D$. The function $\mathtt{build\_quadratic\_form}(t,\xb_j)$ then returns the two quadratic forms $q^- := q_{\xb_j, \lambda^-}$ and $q^+ := q_{\xb_j, \lambda^+}$.

\begin{figure}[t]
\begin{algorithmic}[1]                    
\Require tree $t$, box $K$, SDP relaxation order $k$, control points sequence $s = \{\xb_1,\dots,\xb_r\} \subset K$
\Ensure lower bound $m$,  upper bound $M$, lower semialgebraic estimator $t_2^-$, upper semialgebraic estimator $t_2^+$  
	\If {$t \in \mathcal{A}$}\label{line:begin_ecc}
	\State $t^- := t$, $t^+ := t$	
	%\Return $\minsa$ ($t, \, k$), $\maxsa$ ($t, \, k$), $t$, $t$
	\ElsIf {$\bop$ $ := \mathtt{root}$ $(t)$ is a binary operation with children $c_1$ and $c_2$}
	    \State $m_{c_i}, M_{c_i}, c_{i}^-, c_{i}^+ := \sampapprox{c_i}{K}{k}{s}$ for $i \in \{1,2\}$ 
		\State $t^-, t^+ := \mathtt{compose\_bop} (c_{1}^-, c_{1}^+, c_{2}^-, c_{2}^+)$ \label{line:samp_template1}
	\ElsIf {$r := \mathtt{root} (t) \in \mathcal{T}$ with child $c$}					
		\State $m_c$, $M_c$, $c^-$, $c^+ := \sampapprox{c}{K}{k}{s}$ \label{line:samp_template}
		\State $\parab^-, \parab^+:= \mathtt{build\_par} (r, m_c, M_c, s)$
		\State $t^-, t^+ := \mathtt{compose}(\parab^-, \parab^+, c^-, c^+)$				
		%\State $t_2^-, t_2^+ := \buildtemplate {t}{K}{k}{s}{t_1^-}{t_1^+}$ \label{line:build_template1}
		%\State \Return $\minsa$ ($t_2^-, \, k$), $\maxsa$ ($t_2^+, \, k$), $t_2^-$, $t_2^+$
		%\State $t_2^-, t_2^+ := \buildtemplate {t}{K}{k}{s}{t_1^-}{t_1^+}$ \label{line:build_template2}
		%\State \Return $\minsa(t_2^-, k)$, $\maxsa(t_2^+, k)$, $t_2^-$, $t_2^+$	 
	\EndIf\label{line:end_ecc}
	\State $t_2^-, t_2^+ := \buildtemplate {t}{K}{k}{s}{t^-}{t^+}$
	\State \Return $\minsa(t_2^-, k)$, $\maxsa(t_2^+, k)$, $t_2^-$, $t_2^+$	
\end{algorithmic}
\caption{$\templateoptim$}
\label{alg:samp_template}
\end{figure}

\begin{figure}[t]
\begin{algorithmic}[1]                    
\Require tree $t$, box $K$, SDP relaxation order $k$, control points sequence $s = \{\xb_1,\dots,\xb_r\} \subset K$, lower/upper semialgebraic estimator $t^-$, $t^+$
%\Ensure  lower semialgebraic estimator $t_2^-$, upper semialgebraic estimator $t_2^+$
%	\If {$t$ is marked with the \textit{template} label} 
    \If {the number of lifting variables exceeds $\lift^{\max}$}
\label{line:cnd_template}
	\For {$\xb_j \in s$} 
		\State $q_j^-, q_j^+ := \tmalgo {t}{\xb_j}$ \label{line:tmalgo}
		\State $m_j^- := \minsa(t_1^- - q_j^-, k)$ \Comment{$ q_j^- + m_j^- \leq t^- \leq t $}\label{line:new_sa}
		\State $M_j^+ := \maxsa(q_j^+ - t_1^+, k)$ \Comment{$ q_j^+ + M_j^+ \geq t^+ \geq t $}
	\EndFor					
	\State \Return $\max_{1 \leq j \leq r} \{q_j^- + m_j^-\}$, $\min_{1 \leq j \leq r} \{q_j^+ + M_j^+\}$
	\Else
		\State \Return $t^-$, $t^+$
	\EndIf
\end{algorithmic}
\caption{$\tt{build\_template}$}
\label{alg:build_template}
%\end{minipage}
%\caption{Our non-linear templates method $\tt{samp\_template}$, based
%  on a recursive semialgebraic max-plus approximation algorithm}
%\label{alg:algo_samp}
\end{figure}

\begin{example}[Modified Schwefel Problem]
\label{ex:swf}
We illustrate our method with the function $f$ from
Example~\ref{ex:sin} and the finite set of three control points
$\{135, 251, 500\}$. For each $i=1,\dots,n$, consider the sub-tree
$\sin(\sqrt{x_i})$. First, we represent each sub-tree $\sqrt{x_i}$ by
a lifting variable $y_i$ and compute $a_1 := \sqrt{135}$, $a_2 :=
\sqrt{251}$, $a_3 := \sqrt{500}$. Then, we get the equations of
$\parab_{a_1}^-$, $\parab_{a_2}^-$ and $\parab_{a_3}^-$ with
$\tt{build_{par}}$, which are three underestimators of the 
function $\sin$ on the real interval $I := [1, \sqrt {500}]$. Similarly we
obtain three overestimators $\parab_{a_1}^+$, $\parab_{a_2}^+$ and
$\parab_{a_3}^+$. Finally, we obtain the underestimator $t_{1, i}^- :=
\max_{j \in \{1, 2, 3\}} \{ \parab_{a_j}^- (y_i)\}$ and the
overestimator $t_{1, i}^+ := \min_{j \in \{1, 2, 3\}}
\{ \parab_{a_j}^+ (y_i)\}$. To solve the modified Schwefel problem, we
consider the following POP:
\[
	\left\{			
	  \begin{array}{ll}
		\min\limits_{\xb \in [1, 500]^n, \yb \in [1, \sqrt{500}]^n, \zb \in [-1, 1]^n} 
          & - \sum_{i=1}^n (x_i + \epsilon x_{i + 1}) z_i \\			 
	       \text{s.t.} 
           & z_i \leq \parab_{a_j}^+(y_i), j \in \{1, 2, 3\},  i = 1,\cdots,n\\
          & y_i^2 = x_i, i = 1,\cdots,n \\
	\end{array} \right.
\]

Notice that the number of lifting variables is $2 n$ and the number of
equality constraints is $n$, thus we can obtain coarser semialgebraic
approximations of $f$ by considering the function $b \mapsto
\sin(\sqrt{b})$ (see Figure~\ref{fig:sin_sqrt3}). We get new
estimators $t_{2, i}^-$ and $t_{2, i}^+$ of each sub-tree
$\sin(\sqrt{x_i})$ with the functions
$\mathtt{build\_quadratic\_form}$, $\minsa$ and
$\maxsa$. The resulting POP involves only $n$ lifting
variables. Besides, it does not contain equality constraints anymore, which improves in practice the numerical stability of the POP solver.

\if{
\begin{figure}[!ht]
\begin{center}
\begin{tikzpicture}[xscale=0.4]	
		
\draw[->] (0,0) -- (27,0) node[right] {$a$};
\draw[->] (0,-1.5) -- (0,1.5) node[above] {$y$};
			
\draw[color=whitegreen, dotted, thick] plot [domain=11:13]  (\x, {-0.5*(\x - 11.6)^2 + 0.551* (\x -11.6) - 0.834}) node[anchor = north east] {$\parab_{a_1}^-$};
\draw[color=whitegreen, dotted, thick] plot [domain=13:16.5]  (\x, {-0.5*(\x - 15.83)^2 - 0.992* (\x - 15.83)  -0.12 }) node[anchor = north west] {$\parab_{a_2}^-$};
\draw[color=whitegreen, dotted, thick] plot [domain=20:23]  (\x, {-0.5*(\x -22.36)^2 - 0.933* (\x -22.36)  -0.361 }) node[anchor = south west] {$\parab_{a_3}^-$};

\draw[color=darkgreen, dotted, thick] plot [domain=9:13]  (\x, {0.5*(\x - 11.6)^2 + 0.551* (\x -11.6) - 0.834}) node[anchor = north east] {$\parab_{a_1}^+$};
\draw[color=darkgreen, dotted, thick] plot [domain=15:18.5]  (\x, {0.5*(\x - 15.83)^2 - 0.992* (\x - 15.83)  - 0.12 }) node[anchor = north west] {$\parab_{a_2}^+$};
\draw[color=darkgreen, dotted, thick] plot [domain=21.5:25]  (\x, {0.5*(\x -22.36)^2 - 0.933* (\x -22.36)  - 0.361 }) node[anchor = south west] {$\parab_{a_3}^+$};

\draw[color=red!80, domain=1:22.36, samples=100] plot (\x,{sin(\x r)}) node[anchor = north east] {$ $};

\draw [color=red!80] (5,0.5) node {$\sin$};
\draw (1,-0.3) node {$1$}; %\draw (23,-0.3) node {$M$}; 
%\draw (3,3pt) -- (3,-3pt); 
\draw (22.36,3pt) -- (22.36,-3pt);

\draw (11.6,-0.3) node {$a_1$}; \draw (11.6,3pt) -- (11.6,-3pt); \draw [black, dashed] (11.6,0) -- (11.6,- 0.834);
\draw (15.83,-0.3) node {$a_2$}; \draw (15.83,3pt) -- (15.83,-3pt);
\draw (22.36,-0.3) node {$a_3 = \sqrt{500}$};  \draw (22.36,3pt) -- (22.36,-3pt); \draw (1,3pt) -- (1,-3pt);
\end{tikzpicture}
\caption{A hierarchy of Max-plus Semialgebraic Estimators for $\sin$:\\ $t_{1, i}^- := \max_{j \in \{1, 2, 3\}} \{ \parab_{a_j}^- (\sqrt{x_i})\} \leq  \sin{\sqrt{x_i}} \leq t_{1, i}^+ := \min_{j \in \{1, 2, 3\}} \{ \parab_{a_j}^+ (\sqrt{x_i})\} $}	\label{fig:sin3}
\end{center}
\end{figure}
}\fi

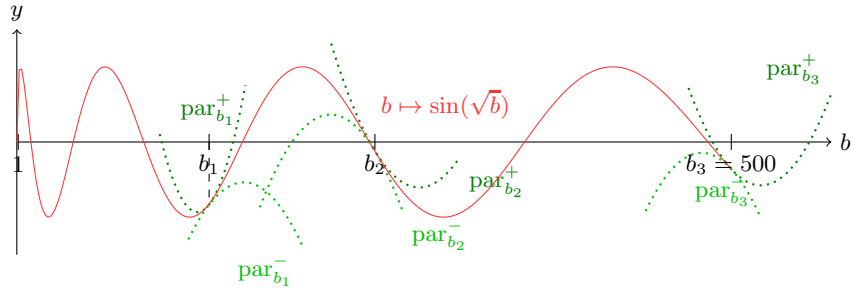
\begin{figure}[t]
\begin{center}
\begin{tikzpicture}[xscale=0.019]	
		
\draw[->] (1,0) -- (570,0) node[right] {$b$};
\draw[->] (0,-1.5) -- (0,1.5) node[above] {$y$};
\draw [color=red!80] (300,0.5) node {$ b \mapsto \sin (\sqrt{b}) $};	
\draw[color=whitegreen, dotted, thick] plot [domain=120:200]  (\x, {-0.0005*(\x - 134.56)^2 + 0.0245* (\x - 134.56) - 0.834}) node[anchor = north east] {$\parab_{b_1}^-$};
\draw[color=whitegreen, dotted, thick] plot [domain=170:270]  (\x, {-0.0005*(\x - 250.6)^2  -0.0313* (\x - 250.6)  -0.12 }) node[anchor = north west] {$\parab_{b_2}^-$};
\draw[color=whitegreen, dotted, thick] plot [domain=440:520]  (\x, {-0.0005*(\x -500)^2 -0.0209* (\x -500)  -0.361 }) 
node[anchor = south east] {$\parab_{b_3}^-$};

\draw[color=darkgreen, dotted, thick] plot [domain=100:160]  (\x, {0.0015*(\x - 134.56)^2 + 0.0245* (\x - 134.56) - 0.834}) node[anchor = north east] {$\parab_{b_1}^+$};
\draw[color=darkgreen, dotted, thick] plot [domain=220:310]  (\x, {0.0005*(\x - 250.6)^2  -0.0313* (\x - 250.6)  -0.12 }) node[anchor = north west] {$\parab_{b_2}^+$};
\draw[color=darkgreen, dotted, thick] plot [domain=470:570]  (\x, {0.0005*(\x -500)^2 -0.0209* (\x -500)  -0.361 }) 
node[anchor = south east] {$\parab_{b_3}^+$};

\draw[color=red!80, domain=0:500, samples=300] plot (\x,{sin(sqrt(\x) r)}) node[anchor = south east] {$ $};

\draw (1,-0.3) node {$1$};
\draw (134.56,-0.3) node {$b_1$}; \draw (134.56,3pt) -- (134.56,-3pt); \draw [black, dashed] (134.56,0) -- (134.56,- 0.834);
\draw (250.6,-0.3) node {$b_2$};\draw (250.6,3pt) -- (250.6,-3pt);
\draw (500,-0.3) node {$b_3 = 500$};  \draw (500,3pt) -- (500,-3pt); \draw (1,3pt) -- (1,-3pt);
%\draw (1,-0.3) node {$m$}; \draw (25.5,-0.3) node {$M$}; 
%\draw (3,3pt) -- (3,-3pt); 
%\draw (25,3pt) -- (25,-3pt);

%\draw (11.6,-0.3) node {$a_1$};  \draw (11.6,3pt) -- (11.6,-3pt); \draw [black, dashed] (11.6,0) -- (11.6,- 0.834);
%\draw (15.83,-0.3) node {$a_2$};  \draw (-1,3pt) -- (-1,-3pt); \draw [black, dashed] (-1,0) -- (-1,- 0.12);
%\draw (22.36,-0.3) node {$a_3$};  \draw (1,3pt) -- (1,-3pt); \draw [black, dashed] (1,0) -- (1,- 0.361);
\end{tikzpicture}
\caption{Templates based on Max-plus Semialgebraic Estimators for $b \mapsto \sin(\sqrt{b})$:\\ $t_{2, i}^- := \max_{j \in \{1, 2, 3\}} \{ \parab_{b_j}^- (x_i)\} \leq  \sin{\sqrt{x_i}} \leq t_{2, i}^+ := \min_{j \in \{1, 2, 3\}} \{ \parab_{b_j}^+ (x_i)\} $}\label{fig:sin_sqrt3}
%\caption{A hierarchy of Semialgebraic Estimators for $b \mapsto \sin(\sqrt{b})$}	\label{fig:sin_sqrt3}
\end{center}
\end{figure}
\end{example}

\paragraph{Dynamic choice of the control points.}

As in~\cite{victorecc}, the sequence $s$ of control points is computed iteratively. We initialize the set $s$ to a single point of $K$, chosen so as to be a minimizer candidate for $t$ (e.g. with a local optimization solver). Calling the algorithm $\templateoptim$ on the main objective function $t_f$ yields an underestimator $t_f^-$. Then, we compute a minimizer candidate $\Xopt$ of the underestimator tree $t^-_f$. It is obtained by projecting a
solution $\Xsdp$ of the SDP relaxation of Section~\ref{sect:sa} on the
coordinates representing the first order moments,
following~\cite[Theorem~4.2]{DBLP:journals/siamjo/Lasserre01}. We add $\Xopt$ to the set of control points $s$. Consequently, we can refine dynamically our templates based max-plus approximations by iterating the previous procedure to get tighter lower bounds. This procedure can be stopped as soon as the requested lower bound is attained.

\begin{remark}[Exploiting the system properties]
  Several properties of the POP can be exploited to
  decrease the size of the SDP relaxations such as
  symmetries~\cite{DBLP:journals/corr/abs-1103-0486} or
  sparsity~\cite{DBLP:journals/toms/WakiKKMS08}.
  Consider Problem~\eqref{eq:f} with $f$ having some sparsity pattern
  or being invariant under the action of a finite subgroup
  symmetries. Then the same properties hold for the resulting
  semialgebraic relaxations that we build with our non-linear
  templates method.
\end{remark}

\section{RESULTS}
\label{sect:bench}

\subsubsection{Comparing three certification methods.}
We next present numerical results 
obtained by applying the present template method
to examples from the global optimization literature,
as well as inequalities from the Flyspeck project.
Our tool is implemented in OCaml and interfaced with the SparsePOP solver~\cite{DBLP:journals/toms/WakiKKMS08}.

In each example, our aim is to certify a lower bound $m$ of a function $f$ on a box $K$. 
We use the algorithm $\templateoptim$,
keeping the SOS relaxation order $k$ sufficiently small
to ensure the fast computation of the lower bounds. 
The algorithm $\templateoptim$ returns more precise bounds by successive updates of the control points sequence $s$. However, in some examples,
the relaxation gap is too high to certify the requested
bound. 
Then, we perform a domain subdivision in order to reduce this gap:
we divide the maximal width interval of $K$ in two halves to get two sub-boxes $K_1$ and $K_2$ such that $K = K_1 \cup K_2$. We repeat this subdivision procedure,  by applying $\templateoptim$ on a finite set of sub-boxes,
until we succeed to certify that $m$ is a lower bound of $f$.
We denote by $\nbb$ the total number of sub-boxes generated
by the algorithm. 

For the sake of comparison, we have implemented a template-free SOS method $\iasos$, which coincides with the particular case of $\templateoptim$ in which $\#s = 0$ and $\lift = 0$.
It computes the bounds of semialgebraic functions with standard SOS relaxations and bounds the univariate transcendental functions by interval arithmetic. 
We also tested the MATLAB toolbox algorithm $\tt{intsolver}$~\cite{Montanher09}, which is based on the Newton interval method~\cite{Hansen198389}. Experiments are performed on an Intel Core i5 CPU ($2.40\, $GHz).

\subsubsection{Global optimization problems.}
\label{subsect:bench1}
 
The following test examples are taken from Appendix B in~\cite{Ali:2005:NES:1071327.1071336}. Some of these examples depend on numerical constants, the values of which can be found there. 

%\if{
\begin{itemize}
\item {\em Hartman 3 (H3)}: $\min\limits_{\xb \in [0, 1]^3} f(\xb) = - \sum\limits_{i=1}^4 c_i \exp \left[- \sum\limits_{j=1}^3 a_{i j} (x_j - p_{i j})^2\right] $
%\item {\em Hartman 3}
\item {\em Mc Cormick (MC)}, with $K = [-1.5, 4] \times [-3, 3] $:\\ 
$\min\limits_{\xb \in K} f(\xb) = \sin (x_1 + x_2) + (x_1 - x_2)^2 - 0.5 x_1 + 2.5 x_2 + 1$
\item {\em Modified Langerman (ML)}:\\
$\min\limits_{\xb \in [0, 10]^n} f(\xb) = \sum\limits_{j=1}^5 c_j \cos (d_j / \pi) \exp (- \pi d_j) $, with $d_j = \sum\limits_{i=1}^n (x_i - a_{j i})^2$
\item {\em Paviani Problem (PP)}, with $K = [2.01, 9.99]^{10}$:\\
$\min\limits_{\xb \in K} f(\xb) = \sum\limits_{i=1}^{10} \left[ (\log(x_i - 2))^2 - \log(10 - x_i))^2 \right]  - \left(\prod\limits_{i=1}^{10} x_i \right)^{0.2}$
\item {\em Shubert (SBT)}:$\min\limits_{\xb \in [-10, 10]^n} f(\xb) = \prod\limits_{i=1}^{n} \Big( \sum\limits_{j=1}^5 j \cos ((j+1) x_i + j) \Big)$
\item {\em Modified Schwefel (SWF)}: see Example~\ref{ex:sin}
\end{itemize}
%}\fi

\paragraph{Informal certification of lower bounds of non-linear problems.}
In Table~\ref{table:go}, the {\em time} column indicates the total informal verification time, \ie{} without the exact certification of the lower bound $m$ with Coq. Each occurrence of the symbol ``$-$'' means that $m$ could not be determined within one day of computation by the corresponding solver.
We see that $\iasos$ already outperforms the interval arithmetic solver $\tt{intsolver}$ on these examples. However, it can only be used for problems with a moderate
number of variables. The algorithm $\templateoptim$ allows us to overcome this restriction, while keeping a similar performance (or occasionally improving this performance) on moderate size examples. 
 
Notice that reducing the number of lifting variables
allows us to provide more quickly coarse bounds for large-scale instances of {\em SWF}. We discuss the results appearing in the two last lines of Table~\ref{table:go}.
Without any box subdivision, we can certify a better lower bound $m = -967 n$ with $\lift = 2n$ since our semialgebraic estimator is more precise. However the last lower bound $m = -968 n$ can be computed twice faster by considering only $n$ lifting variables, thus reducing the size of the POP described in Example~\ref{ex:swf}. This indicates that the method is able to avoid the blow up for certain hard sub-classes of problems where a standard
(template free) POP formulation would involve a large number of lifting variables.  

\paragraph{Formal certification of lower bounds of POP.} For some
small size instances of POP, our tool can prove the correctness of
lower bounds. Our solver is interfaced with the framework mentioned
in~\cite{Monniaux_Corbineau_ITP2011} to provide exact rational
certificates, which can be formally checked with Coq. This formal
verification is much slower. As an example, for the {\em MC} problem,
it is $36$ times slower to generate exact SOS certificates and $13$
times slower to prove its correctness in Coq. Note that the interface
with Coq still needs some streamlining.
%Pour certaines inegalites avec peu de variables, on obtient des bornes certifiees en Coq pour les POP. Par exemple, la verification formelle de MC est 50 fois plus lente (2\% du temps en verif informelle, 73\% en projection et 25\% en normalisation en Coq)

\paragraph{High-degree polynomial approximations.}
An alternative approach consists in approximating the transcendental functions by polynomial functions of sufficiently high degree, and then applying sums of squares approach to the polynomial problems. Given $d \in \N$ and a floating-point interval $I$, we can approximate an univariate transcendental function on $I$ by the best uniform degree-$d$ polynomial approximation and obtain an upper bound of the approximation error. This technique, based on Remez algorithm, is implemented in the Sollya tool (for further details, see e.g.~\cite{Brisebarre:2010:CIP:1837934.1837966}).

We interfaced our tool with Sollya and performed some numerical
tests. The minimax approximation based method is eventually faster
than the templates method for moderate instances. For the examples
{\em H3} and {\em H6}, the speed-up factor is $8$ when the function
$\exp$ is approximated by a quartic minimax polynomial.

However, this approach is much slower to compute lower bounds of problems involving a large number of variables. It requires $57$ times more CPU time to solve {\em SWF ($\epsilon = 1$)} with $n = 10$ by considering a cubic minimax polynomial approximation of the function $b \mapsto \sin(\sqrt{b})$ on a floating-point interval $I \supseteq [1, \sqrt{500}]$. These experiments indicate that a high-degree polynomial approximation is not suitable for large-scale problems.

\begin{table}[!t]
\begin{center}
\caption{Comparison results for global optimization examples}
\begin{tabular}{|lcc|ccccr|cr|r|}
\hline
\multirow{2}{*}{Problem}
 &
\multirow{2}{*}{$n$}
 & 
\multirow{2}{*}{$m$}
&
\multicolumn{5}{c|}{$\templateoptim$}&
\multicolumn{2}{c|}{$\iasos$} & 
\multicolumn{1}{c|}{$\tt{intsolver}$}
\\
\cline{4-11}
& & & $k$ & $\#s$ & $\lift$ & $\nbb$ & time & $\nbb$  &time  & time\\
%\hline
\hline            
{\em H3} & $3$ &  $ -3.863 $ &  $2 $ & $3$ &  $4$ & $99$  & $101 \, s$ & $1096$ & $247 \, s$ &  $3.73 \, h$\\
%\hline
{\em H6} & $6$ &  $ -3.33 $ &  $2 $ & $1$ &  $6$ & $113$  & $102 \, s$ & $113$ & $45 \, s$ &  $> 4 \, h$\\
%\hline
{\em MC} & $2$ &  $ -1.92 $ &  $1 $ & $2$ &  $1$ & $17$  & $1.8 \, s$ & $92$ & $7.6 \, s$ & $4.4 \, s$\\
%\hline
{\em ML} & $10$ &  $ -0.966 $ &  $1$ & $1$ &  $6$ & $8$  & $8.2 \, s$ & $8$ & $6.6 \, s$ & $> 4 \, h$\\
%\hline
{\em PP} & $10$ &  $ -46 $ &  $1$ & $3$ &  $2$ & $135$  & $89 \, s$ & $3133$ & $115 \, s$ & $56 \, min $\\
%\hline
{\em SBT} & $2$ &  $ -190 $ &  $2$ & $3$ &  $2$ & $150$  & $36 \, s$ & $258$ & $0.6 \, s$ & $57 \, s$\\
%\hline
\hline
\multirow{4}{*}{{\em SWF ($\epsilon = 0$)}} & $10$ &  $ -430 n$ &  $2$ & $6$ &  $2 n$ & $16$  & $40 \, s$ & $3830$ & $129 \, s$ & $18.5 \, min$\\
%\cline{2-11}
 & $100$ &  $ -440 n$ &  $2$ & $6$ &  $2 n$ & $274$  & $1.9 \, h$ & $> 20000$ & $> 10 \, h$ & $-$\\
%\cline{2-11}
& $1000$ &  $ -486 n$ &  $2$ & $4$ &  $2 n$ & $1$  & $450 \, s$ & $-$ & $-$ & $-$\\
%\cline{2-11}
 & $1000$ &  $ -488 n$ &  $2$ & $4$ &  $n$ & $1$  & $250 \, s$ & $-$ & $-$ & $-$\\
%\hline
\hline
\multirow{2}{*}{{\em SWF ($\epsilon = 1$)}} & $1000$ &  $ -967 n$ &  $3$ & $2$ &  $2 n$ & $1$  & $543 \, s$ & $-$ & $-$ & $-$\\
%\cline{2-11}
 & $1000$ &  $ -968 n$ &  $3$ & $2$ &  $n$ & $1$  & $272 \, s$ & $-$ & $-$ & $-$\\
\hline
\end{tabular}
\label{table:go}
\end{center}
\end{table}

\begin{table}[!t]
\begin{center}
\caption{Results for Flyspeck inequalities using $\templateoptim$ with $n = 6$, $k = 2$ and $m = 0$
}
\begin{tabular}{|l|rrccr|}
\hline
Inequality id & $\nt$   & $\#s$ & $\lift$   & $\nbb$ & time\\
\hline
%\hline			
 $9922699028$ & $1$   &$4$ & $9$ & $47$   &   $241 \, s$ \\
 $9922699028$ & $1$   &$4$ & $3$ & $39$   &   $190 \, s$ \\
%\hline			
 $3318775219$ & $1$   &$2$ & $9$& $338$   &   $26 \, min$ \\
%\hline	
 $7726998381$ & $3$   & $4$ & $15$&  $70$  & $43 \, min$ \\
%\hline
 $7394240696$ & $3$    & $2$ & $15$& $351$  & $1.8 \, h$\\
% \hline
 $4652969746\_1$ & $6$    &$4$ & $15$ & $81$  & $1.3 \, h$\\
 %\hline
 $\text{OXLZLEZ}\, 6346351218\_2\_0$ &  $6$ & $4$& $24$  & $200$  & $5.7 \, h$\\
\hline						
\end{tabular}	
%$1.1 \times 10^{-5}$
%$2.0 \times 10^{-5}$
%$1.4 \times 10^{-5}$
%$1.9 \times 10^{-5}$
%$9.3 \times 10^{-6}$
%$2.3 \times 10^{-6}$
\label{table:flyspeck}		
\end{center}	
\end{table}

\subsubsection{Certification of various Flyspeck inequalities.}
\label{subsect:bench2}

In Table~\ref{table:flyspeck}, we present some test results for several non-linear Flyspeck inequalities. The information in the columns {\em time}, $\nbb$, and $\lift$ is the same as above. 
%The column {\em time} contains the informal verification time. 
The integer $\nt$ represents the number of transcendental univariate nodes in the corresponding abstract syntax trees. 
These inequalities are known to be tight and involve sum of arctan of correlated functions in
many variables, whence we keep high the number of lifting variables to get precise max-plus estimators. However, some inequalities (e.g. $9922699028$) are easier to solve by using coarser semialgebraic estimators.
For instance, the first line ($\lift = 9$) corresponds to the algorithm described in~\cite{victorecc} and the second one ($\lift = 3$) illustrates our improved templates method. For the latter, we do not use any lifting variables to represent square roots of univariate functions.

\section{CONCLUSION}
The present quadratic templates method computes certified lower bounds
for global optimization problems. It can provide tight max-plus
semialgebraic estimators to certify non-linear inequalities involving
transcendental multivariate functions (e.g.\ for Flyspeck
inequalities). It also allows one to limit the growth of the number of
lifting variables as well as of polynomial constraints to be handled
in the POP relaxations, at the price of a coarser approximation.
%in our POP relaxations can be bounded with a $\lift$ parameter.
Thus, our method is helpful when the size of optimization problems
increases.
%  increase the size of certifiable
%certify bounds for non-linear moderate-scale 
% optimization problems. 
Indeed, the coarse lower bounds obtained (even
with a low SDP relaxation order) are better than those obtained with
interval arithmetic or high-degree polynomial approximation. For
future work, we plan to study how to obtain more accurate non-linear
templates by constructing a sequence of semialgebraic estimators,
which converges to the ``best'' max-plus estimators (following the
idea of~\cite{DBLP:conf/cdc/LasserreT11}).

Furthermore, the formal part of our implementation, currently can only
handle small size POP certificates. We plan to address this issue by a
more careful implementation on the Coq side, but also by exploiting
system properties of the problem (sparsity, symmetries) in order
to reduce the size of the rational SOS certificates.
Finally, it remains to complete the formal verification procedure
by additionally proving in Coq the correctness of our semialgebraic
estimators.

\section*{Acknowledgements}
The authors thank the anonymous referees for helpful comments and suggestions to improve this paper.

%\bibliography{mybib}

\end{document}